\documentclass[%
aps, nobibnotes ,secnumarabic ,showpacs ,amssymb, amsmath, prb
]{revtex4}

\usepackage{endnotes}

\let\footnote=\endnote

\def\a{\alpha}
\def\be{\begin{equation}}\def\ee{\end{equation}}\def\l{\label}
\def\0{\setcounter{equation}{0}}\def\C{\cite}
\def\r{\ref}\def\ba{\begin{eqnarray}}\def\ea{\end{eqnarray}}
\def\x{\xi}\def\la{\lambda}
\def\d{\delta}\def\s{\sigma}\def\f{\frac}\def\D{\Delta}\def\pa{\partial}
\def\o{\omega}\def\O{\Omega}\def\ga{\gamma}
\def\Ga{\Gamma}\def\t{\times}
\def\vp{\varphi}\def\le{\left}\def\ri{\right}
\def\foot{\footnote}\def\vep{\varepsilon}
\def\k{\kappa}\def\bx{{\mathbf x}}\def\La{\Lambda}

\def\Re{{\rm Re}}\def\Im{{\rm Im}}
\def\bk{\mathbf{k}}\def\cn{{\cal N}}
\def\cg{{\cal G}}\def\q{\eta}
\def\bE{{\mathbf e}}
\def\cg{{\cal G}}\def\ch{{\cal H}}
\def\kb{{\mathbb
K}}\def\cd{\mathcal D}\def\mj{\mathbf{J}}
\def\bu{{\mathbf u}}\def\by{{\mathbf y}}\def\bp{{\mathbf p}}
\def\k{\kappa} \def\cz{{\mathcal Z}}\def\ma{\mathbf{A}}
\def\me{\mathbf{E}}\def\mp{\mathbf{P}}
\def\ru{\mathrm{u}}\def\rp{\mathrm{p}}\def\z{\zeta}

\begin{document}

\title{On invariant quantization of non-Abelian gauge fields}
\author{J.~Manjavidze}\email{joseph@jinr.ru}
\affiliation{Tbilisi State University, Institute of Physics, Tbilisi,
Georgia} \affiliation{Joint Institute for Nuclear Research, Dubna
141980, Russia}

\begin{abstract}
A strong coupling expansion around the non-trivial extremum of the
Yang-Mills action will be described. It is shown that the developed
formalism is the Gribov ambiguity free and each order of the
developed perturbation theory is transparently gauge invariant. The
result is a consequence of the restriction: calculations are not
going beyond the module square of the $S$-matrix elements.
\end{abstract}

\pacs{11.10.Lm, 11.15.-q, 11.15.Kc}

\received{}

\maketitle

\section{Introduction}

We present a partial solution of the Gribov problem \C{gribov,atiyah,
zinger}: it is impossible to extract unambiguously the non-Abelian
gauge symmetry degrees of freedom by Faddeev-Popov {\it ansatz} if
the gauge field is strong, see also \C{faddeev1, shabanov}. At the
same time the canonical quantization scheme certainly prescribes to
extract the symmetry degrees of freedom \C{dirak}. Therefore, the
problem of quantization of the non-Abelian gauge theories is on hand.

Our aim is to show that the problem can be solved if the module
square of the amplitude, $|\mathbb A|^2$, is calculated. The
quantization problem becomes simpler since the phase of $\mathbb A$
is excluded from consideration in this case. In other words, we will
argue that the Yang-Mills quantum field theory is free from Gribov
ambiguities if it is used for the phase-free quantities description.
The application of this restricted formalism was deduced in a number
of papers, see e.g. \C{elpat, physrep}.

It can be shown\C{yaph} that functional integral representation for
$|\mathbb A|^2$ is defined on the $\delta$-like (Dirac) measure: \be
DM(A)=\prod_{x, a}dA_{a\mu}(x)\d\le( \f{\d S (A)}{\d
A^{a\mu}(x)}+\hbar J_{a\mu}(x)\ri),\l{1.1}\ee where $A_{a\mu}$ is the
Yang-Mills potential, $a$ is the color index. It is important that no
gauge fixing procedure is assumed for derivation of (\r{1.1}), see
Sec.2. The Dirac measures appearance is the consequence of
cancelations: the phase of $\mathbb A$ can stay arbitrary when the
measurables, $\sim\mathbb A\mathbb A^\dag$, are calculated, Sec.2.

We will consider in present paper the solution in the frame of which
quantum source $\hbar J_{a\mu}(x)$ is switched on adiabatically, i.e.
we will searching for a solution of the equation: \be \f{\d S (A)}{\d
A^{a\mu}(x)}+\hbar J_{a\mu}(x)=0 \l{1.1b}\ee in the form of power
series over $\hbar J_{a\mu}(x)$. The "generalized correspondence
principle" written in (\r{1.1}) is strict for arbitrary value of
$\hbar$ and, therefore, the functional integral defined on the
measure (\r{1.1}) permits the arbitrary transformations\foot{Notice
that the general transformations of functional integral leads to the
wrong results \C{edwards, marinov}, see also \C{shabanov}. }. The
point is that (\r{1.1}) defines the rule how the quantum force,
$J_{a\mu}(x)$, must be transformed if the field, $A_{a\mu}(x)$, is
transformed.

It is the Dirac measure requires to perform the transformation in the
class of strict solutions, $u_{a\mu}(x)$, of the sourceless (with
$J_{a\mu}=0$) Lagrange equation. This stands for mapping into the
coset space $W$, \C{hioe, isham, linden, landsman, mackey}: \be
u_{a\mu}:~~A_{a\mu}\to \{\la\}\in W.\l{1.2}\ee Here $W$ corresponds
to the factor group $\cg/\ch$, where $\cg$ is the symmetry group of
the problem and $\ch$ is the invariance group of $u_{a\mu}$. The
qualitative reason of this choice is following: after having got the
ground state field $u_{a\mu}(x)$, where $u_{a\mu}(x)$ is any strict
solution of sourceless Lagrange equation, the freedom in the choice
of the value of integration constants, $\{\la\}$, is what remains
from the continuum of the field degrees of freedom, see Sec.3. The
gauge phases $\{\La^a\}$ must be included in $\{\la\}$.

It should be noted that the mapping (\r{1.2}) may be singular if
$\dim W$ is finite. We will show that the singularity can be isolated
(and canceled by the normalization). This is our renormalization
procedure. Sec.~\r{re1} contains description of this procedure.

It will be shown that each order of our perturbation series is
transparently gauge invariant since the gauge-invariant, $\mathbb
A{\mathbb A}^\dag$, will be calculated. This will be shown in
Sec.~\r{g:inv}. Therefore, no gauge fixing procedure is required and
no ambiguities appears. This is the main result. The preliminary
version of the formalism was given in \C{jmp-3}.

\section{Perturbation theory}

We will consider the theory with the action:\be
S(A)=-\f{1}{4g^2}\int d^4x\: F_{\mu\nu}^a(A)F^{\mu\nu}_a(A)
\l{f3}\ee The Yang-Mills fields \be F_{a\mu\nu}(A)=\pa_\mu
A_{a\nu}-\pa_\nu A_{a\mu}- f_{abc}A_{b\mu}A_{c\nu} \l{f4}\ee are
the non-Abelian gauge covariants. The group will not be specified.
The matrix notation: $A_{a\mu}\o_a=A_{\mu}$ will be also used.

We will calculate the quantity\foot{The generalization was considered
in \C{physrep, tmf}.}: \be \cn=|\cz|^2,\l{}\ee where the
vacuum-into-vacuum transition amplitude \be \cz=\int DA\:
e^{iS_C(A)},\quad DA=\prod_{\bx,t\in C}\prod_{a,\mu}\f{d
A_{a,\mu}(\bx,t)}{\sqrt{2\pi}},\l{}\ee is defined on the Minkowski
metric. The Mills complex time formalism will be used to avoid the
possible light cone singularities \C{milss}. For example, the theory
may be defined on the complex time contour \be C:~t\to
t+i\vep,~\vep\to+0,~-\infty\leq t\leq+\infty.\l{}\ee At the very end
one must take $\vep=+0$. The Mills formalism restores the Feynman's
$i\vep$-prescription.

\subsection{\it Dirac measure}

The double integral: \be \cn=\int DA^+DA^-e^{iS_C(A^+)-
iS^*_C(A^-)}\l{2.12}\ee will be calculated. To extract the Dirac
measure\foot{The term "$\d$-like (Dirac) measure" have been taken
from \C{fedor}.}, one can introduce the mean trajectory, $A_{a\mu}$,
and the virtual deviation, $a_{a\mu}$, instead of $A_\mu^\pm$: \be
A_{a\mu} ^\pm(x)=A_{a\mu}(x)\pm a_{a\mu}(x). \l{2.13}\ee The
transformation (\r{2.13}) is linear and the differential measure \be
DA^+DA^- =\prod_{\bx,t\in C+C^* }\prod_{a,\mu}dA_{a\mu}(x)
\prod_{\bx,t\in C+C^*}\prod_{a,\mu}\f{da_{a\mu}(x)} {\pi} \equiv
DADa.\l{2.15a} \ee is defined on the entire time contour $C+C^*$.

The "closed-path" boundary conditions: \be a_{a\mu}(x\in\s_\infty)
=0,\l{2.16a}\ee where $\s_\infty$ is the remote time-like
hypersurface, is assumed. We will demand that the surface terms
are cancelled in the difference $S_C(A^+)-S^*_C(A^-)$, i.e. \be
\int dx \pa_\mu(A_{a\nu}\pa^\mu A^{a\nu})^+=\int dx
\pa_\mu(A_{a\nu}\pa^\mu A^{a\nu})^-\l{}\ee if (\r{2.16a}) is taken
into account. Therefore, not only the trivial pure gauge fields
can be considered on $\s_\infty$.

Expanding $S(A\pm a)$ over $a_{a\mu}$, one can write: \be
S_C(A+a)-S_C^*(A-a)=U(A,a) +2\Re\int_C dx\: a_{a\mu}(x)\f{\d S(A)}
{\d A_{a\mu}(x)}.\l{2.17a} \ee This equality will be used as the
definition of the remaining term, $U(A,a)$. With the $\vep$ accuracy,
$U(A,a)=O(a^3)$, i.e. $U(A,a)$ introduces the interactions.

Noticing that \be \f{\d S(A)} {\d A^{a\mu}(x)}=D^{\nu b
}_aF^b_{\mu\nu},\l{2.19a}\ee where $D^{\nu b }_a$ is the covariant
derivative and inserting (\r{2.15a}) and (\r{2.17a}) into (\r{2.12}),
we find: \be \cn=\int DA\int Da~e^{2i\Re\int_C dx a_a^\mu(x)D^{\nu b
}_aF^b_{\mu\nu}}~e^{iU(A,a)}.\l{2.20a}\ee The integrals over
$a_{a\mu}(x)$ will be calculated perturbatively. For this purpose one
can use the identity: \be
e^{iU(A,a)}=\lim_{\z_{a\mu}=J_{a\mu}=0}e^{-i\:\kb(J,\z)}
e^{2i\:\Re\int_C dx\: a_{a\mu}(x) J^{a\mu}(x)}e^{i\:U(A,\z)},\l{2.21}
\ee where \be 2\kb(J,\z)= \Re\int_C dx\f{\d}{\d J^{a\mu}(x)}\f{\d}{\d
\z_{a\mu}(x)}.\l{2.22a} \ee In the future we will omit  the symbol of
the limit appearing in (\r{2.21}) keeping in mind the prescription:
the auxiliary variables, $J_{a\mu}$ and $\z_{a\mu}$, must be taken
equal to zero at the very end of calculations.

Assuming that the perturbation series will exist, the insertion of
the Eq.(\r{2.21}) into (\r{2.20a}) gives the desired expression:
\be \cn=e^{-i\kb(J,\z)}\int DM(A)e^{iU(A,\z)},\l{2.23}\ee where
\begin{multline}
DM(A)=\prod_{\bx,t\in C+C^*}\prod_{a,\mu}dA_{a\mu}(x)
\int\prod_{\bx,t \in C+C^*} \prod_{a,\mu}\f{da_{a\mu}(x)}
{\pi}e^{2i\Re\int_c dx a^{a\mu}(D^{\nu b
}_aF^b_{\mu\nu}-J_{a\mu}(x))}= \\ = \prod_{\bx,t\in
C+C^*}\prod_{a,\mu}dA_{a\mu}(x)\d\le(D^{\nu b
}_aF^b_{\mu\nu}-J_{a\mu}(x)\ri)\l{2.24}
\end{multline}
is the functional Dirac measure. The functional $\d$-function on the
complex time contour $C+C^*$ has the definition: \be \prod_{\bx,t\in
C+C^*} \d(D^{\nu b }_aF^b_{\mu\nu}-J_{a\mu}) =\prod_{\bx,t\in
C}\d(\Re(D^{\nu b }_aF^b_{\mu\nu}-J_{a\mu})) \d(i\Im(D^{\nu b
}_aF^b_{\mu\nu}- J_{a\mu}))\l{2.26}.\ee It is important that the
exponent in (\r{2.24}) is pure imaginari as the consequence of the
fact that the module $|\cz|^2$ is calculated.

It can be shown that (\r{2.23}) gives the ordinary perturbation
theory (pQCD) \C{jmp-1,jackiw-1, jackiw-2, faddeev} if the
equation: \be D^\nu_{ab}F^b_{\mu\nu}= J_{a\mu} \l{2.25}\ee is
expanded in the vicinity of $A_{a\mu}=0$. Notice that the
Eq.(\r{2.25}) is not gauge invariant because of $J_{a\mu}(x)$.

\section{Mapping into the coset space}

We will formulate the general method of mapping (\r{1.2}) into the
infinite dimensional phase space $\Ga_\infty$, Sec.3.2, and then will
find the reduction procedure, $\Ga_\infty\mapsto W$ on the second
stage of the calculation, Sec.4.1.

\subsection{\it First order formalism}

The action in terms of the electric field, $P^i_a=F^{i0}_a,\:
i=1,2,3,$ looks as follows:
\be
S(\ma,\me)=\f{1}{g^2}\int dx
\le\{\mathbf{P}_a\dot{\ma}_a -
\f{1}{2}(\mathbf{P}_a^2+\mathbf{B}_a^2) +
A_{0a}(\cd\mathbf{P})_a\ri\},\l{p2}
\ee
where the magnetic field $\mathbf{B}_{a}(\ma)= {rot}\ma_{a}- \f{1}{2}(\ma\t \ma)_a$ and
$(\cd\mathbf{P})_a=\pa_i \mathrm{P}_{a i}-f_{abc} \mathrm{A}_{b i}
\mathrm{P}_{c i}$. The
corresponding Dirac measure is:
\be
DM(\ma,\mp)=\prod_{a}\prod_{x}
d\ma_{a}(x)~d\mp_{a}(x)\:\d(\cd\mp_a) \d\le(\dot\mp_a(x)+
\f{\d H_J(\ma,\mp)}{\d\ma_a(x)}\ri) \d\le(\dot\ma_a(x)-\f{\d
H_J(\ma,\mp)}{\d\mp_a(x)}\ri), \l{f10a}
\ee
where
$d{\mathbf{A}}_a(x)d{\mathbf{P}}_{a}(x)=\prod_i d\mathrm{A}_{ia}(x)
d{\mathrm{P}}_{ai}(x),~i=1,2,3,$  and the total Hamiltonian
\be%
H_J=\f{1}{2}\int d^3x\: ({\mathbf P}^2_a+{\mathbf B}_a^2) -\int
d^3x\: {\mathbf J}_a {\mathbf A}_a.\l{a4}\ee Notice that the
dependence on $A_{a0}$ was integrated out and as a result the Gauss
law, $\cd^a_b{\mathbf P}_b=0,$ was appeared in (\r{f10a}). The
Faddeev-Popov $ansatz$ was not used for the definition of the
integral over $A_{a\mu}$. The perturbations generating operator $\kb$
and the remainder potential term $U$ stay unchanged, see (\r{2.22a})
and (\r{2.17a}).

The integrals with the measure (\r{f10a}) will be calculated using
new "collective-like" variables. The same was proposed by Faddeev
and authors\C{faddeev1}. But they introduce the condition
$\cd^a_b{\mathbf P}_b=0$ by hands and their transformation to the
new variables leads to the complicated singular Hamiltonian.

\subsection{\it General mechanism of transformations\l{sec3.2}}

{\bf Proposition 1. } {\it The Jacobian of transformation (\r{1.2})
of the Dirac measure (\r{f10a}) is equal to one} \C{jmp-1}.\\ One can
insert the unite \be
1=\f{1}{\D(\la,\k)}\int\prod_{\a,t}d\la_\a(t)d\k_\a(t)
\prod_{a,\bx}\d(\ma_a(\bx,t)- \bu_a(\bx;\la,\k))\:
\d(\mp_a(\bx,t)-\bp_{a}(\bx;\la,\k)) \l{3.2} \ee into the integral
(\r{2.23}) and integrate over $\ma_a$ and $\mp_a$ using the
$\d$-functions of (\r{3.2}). This is one way to perform the
transformation. Otherwise, if the $\d$-functions of (\r{f10a}) are
used, $\bu_a$ and $\bp_a$ will play the role of constraints. It must
be noted that the both ways of calculation must lead to the identical
ultimate result because of the $\d$-likeness of measures in
(\r{f10a}) and (\r{3.2}). The first way is preferable since it does
not imply the ambiguous gauge fixing procedure \C{gribov, atiyah,
zinger}.

To be correct the power of sets $(\la,\k)$ and $(\ma,\mp)$ must
coincide since in this case only one may introduce transformations
like (\r{1.2}). For this purpose we will consider the theory on the
{\it space} lattice.

The given composite functions $\bu_{a}(\bx;\la(t),\k(t))$ and
$\bp_{a}(\bx; \la(t),\k(t))$ must obey the condition: \be
\D(\la,\k)=\int \prod_{\a,t} d\la'_\a(t) d\k'_\a(t) \prod_{a}
\d(\la'\bu_{a,\la}+\k'\bu_{a,\k}) \d(\la'\bp_{a,\la} +\k'\bp_{a,\k})
\neq0,\l{3.5} \ee where
$$\bu_{a,X}\equiv\f{\pa \bu_{a}}{\pa X}, ~\bp_{a,X}\equiv\f{\pa
\bp_{a}}{\pa X},~~X=(\la_\a,\k_\a)(t)$$ and $(\la,\k)$ are the
solutions of the equations: \be \ma_a(\bx,t)- \bu_a(\bx;\la,\k)=0,~~
\mp_a(\bx,t)-\bp_{a}(\bx;\la,\k)=0. \l{31}\ee The summation over the
repeated index, $\a$, will be assumed. It must be underlined that the
functions $(\bu,\bp)$ are given. Therefore, the equalities (\r{31})
restrict the form of functions $(\ma,\mp)$ on the measure (\r{f10a}).

The transformed measure: \be DM(\la,\k)=\f{1}{\D(\la,\k)} \prod_{\a,
t} d\la_\a(t)d\k_\a(t) \prod_{a}\d\le(\dot{\la}\bu_{a,\la}+
\dot{\k}\: \bu_{a,\k}-\f{\d H_J(\bu,\bp)}{\d
\bp_{a}}\ri)d\le(\dot{\la}\bp_{a,\la}+ \dot{\k}\:\bp_{a,\k}+\f{\d
H_J(\bu,\bp)}{\d \bu_{a}}\ri)\l{32}\ee can be diagonalize introducing
the auxiliary function(al) $h_J$:
\begin{multline}
DM(\la,\k)= \f{1}{\D(\la,\k)}\prod_{\a,t} d\la_\a(t)d\k_\a(t)
\int\prod_{\a,t} d\la_\a'(t) d\k_\a'(t)\\\t
\d\le(\la_{\a}'-\le(\dot{\la}_\a-\frac{\d h_J(\la,\k)}{\d
\k_\a}\ri)\ri) \d\le(\k_\a'-\le(\dot{\k}_\a+\f{\d
h_J(\la,\k)}{\d\la_\a}\ri)\ri) \\ \t\prod_{a}
\d\le(\bu_{a,\la}\la'+\bu_{a,\k}\k'+\{\bu,h_J\}_a -\f{\d H_J}{\d
\bp_a}\ri)\d\le(p_{a,\la} \la'+p_{a,\k}\k'+\{\bp,h_J\}_a+\f{\d
H_J}{\d \bu_a}\ri),\l{3.6}
\end{multline}
where $\{,\}$ is the Poisson bracket and $(\la,\k)$ are the solution
of equations \be \dot{\la}\bu_{a,\la}+ \dot{\k}\: \bu_{a,\k}-\f{\d
H_J(\bu,\bp)}{\d \bp_{a}}=0,~~\dot{\la}\bp_{a,\la}+
\dot{\k}\:\bp_{a,\k}+\f{\d H_J(\bu,\bp)}{\d \bu_{a}}=0, \l{34}\ee see
(\r{32}).

Let us assume now that $\bu_a$, $\bp_a$ and $h_J$ are chosen in such
a way that: \be \{\bu_a,h_J\}-\f{\d H_J}{\d \rp_a}=0,~~\{\bp_a,h_J\}+
\f{\d H_J}{\d \ru_a}=0.\l{3.7}\ee Then, having the condition
(\r{3.5}), the transformed measure takes the form, see (\r{3.6}): \be
DM(\la,\k)= \prod_{\a,t} d\la_\a(\bx,t) d\k_\a(\bx,t)
\:\d\le(\dot{\la}_\a(\bx,t)-\frac{\d h_J(\la,\k)}{\d
\k_\a(\bx,t)}\ri) \d\le(\dot{\k}_\a(\bx,t)+\f{\d
h_J(\la,\k)}{\d\la_\a(\bx,t)}\ri), \l{3.8} \ee where the functional
determinant $\D(\la,\k)$ was canceled since the sets $(\la,\k)$ in
(\r{3.6}) and (\r{3.5}) must coincide if \be
h_J(\la,\k)=H_J(\ru,\rp), \l{3.9}\ee i.e. if $h_J$ is the transformed
"Hamiltonian". One can find the prove in {\bf Proposition 3.}

As a result, \be \cn=e^{-i\kb(J,\z)}\int DM(\la,\k) e^{i
U(\bu,\z)} \l{3.15a}\ee where $\kb(J,\z)$ was defined in
(\r{2.22a}), $DM(\la,\k)$ was defined in (\r{3.8}) and $U(\bu,\z)$
was introduced in (\r{2.17a}). Therefore, the Jacobian of
transformation is equal to one, i.e. in the frame of the
conditions (\r{3.5}) and (\r{3.9}) the phase space volume is
conserved. Q.E.D. \vskip0.3cm

According to (\r{a4}) the transformed hamiltonian $h_J$ is: \be
h_J(\la,\k)=h(\la,\k)-\int d\bx\: \mj_a(\bx,t)
\bu_a(\bx;\la,\k).\l{3.15}\ee Therefore, we come to the following
dynamical problem: \be \dot{\la}_\a=\frac{\d h_J(\la,\k)}{\d
\k_\a}=\frac{\d h(\la,\k)}{\d \k_\a}-\int d\bx\: \mj_a\f{\d\bu_a}{\d
\k_\a} \equiv h_{\k_\a}-\int d\bx\: \mj_a\bu_{a,\k_\a},\l{3.13a} \ee
\be \dot{\k}_\a=-\f{\d h_J(\la,\k)}{\d\la_\a}= -\f{\d
h(\la,\k)}{\d\la_\a}+\int d\bx\: \mj_a\f{\d\bu_a}{\d\la_\a} \equiv
-h_{\la_\a}+\int d\bx\: \mj_a\bu_{a,\la_\a}. \l{3.13b} \ee

\vskip0.25cm{\bf Proposition 2. \it If (\r{3.15}) is held and the
perturbation series exists then the transformation {\rm(\r{1.2})}
induces the splitting: \be \mj_a\to \{j_\la,j_\k\}\l{3.17}\ee} The
proof of the splitting comes from the identity:
\begin{multline}
\prod_{\a,t}\d\le(\dot{\la}_\a-\frac{\d
h_J(\la,\k)}{\d \k_\a}\ri) \d\le(\dot{\k}_\a+\f{\d
h_J(\la,\k)}{\d\la_\a}\ri) =
\exp\{-i\bk(j,e)\}\exp\le\{2i\Re\int_C d\bx dt\mj_a(\bx,t)
(e_{\k_\a}\bu_{a,\la_\a}-e_{\la_\a}\bu_{a,\k_\a})\ri\} \\
\t \prod_{\a,t}
\d(\dot{\la}_\a-h_{\k_\a}-j_{\la_\a}) \d(\dot{\k}_\a+
h_{\la_\a}-j_{\k_\a}),\l{3.14}
\end{multline}
where \be 2\bk(j,e)=\Re\int_C dt dx\le(\f{\d}{\d j_{\la_\a}}\f{\d}{\d
e_{\la_\a}} +\f{\d}{\d j_{\k_\a}}\f{\d}{\d
e_{\k_\a}}\ri).\l{3.15b}\ee At the very end one must take
$j_X=e_X=0$, $X=(\la,\k)$. The equality (\r{3.14}) can be derived
using the functional $\d$-functions Fourier transformation
(\r{2.24}).

Inserting (\r{3.14}) into (\r{3.15a}) and using linearity over
$\mj_a$ of the exponent in (\r{3.14}), we find the completely
transformed representation for $\cn$, where the individual to each
degree of freedom quantum sources, $j_X$, $X=(\la,\k)$, appears. The
transformed representation of $\cn$ looks like: \be\cn=e^{-i\bk(j,e)}
\int DM(\la,\k)e^{iU(\bu,\bE)}, \l{3.20}\ee where \be DM(\la,\k)=
\prod_{\a,\bx, t}d\la_\a(\bx,t) d\k_\a(\bx,t) \:
\d(\dot\la-h_{\k}(\la,k)-j_\la) \d(\dot\k+
h_\la(\la,k)-j_\k),\l{3.18} \ee \be \bE_a=
e_{\k_\a}\bu_{a,\la_\a}-e_{\la_\a}\bu_{a,\k_\a}\l{3.22}\ee and
$\bk(j,e)$ was defined in (\r{3.15b}). Q.E.D.

\vskip0.25cm{\bf Proposition 3.} {\it The Eqs. (\r{3.7}) and the
measure (\r{3.18}) define the classical flow.}\\ Indeed, $$
\dot{\bu}_a= \dot{\la}\bu_{a,\la}+
\dot{\k}\bu_{a,\k}=\{\bu_a,h_J\}=\f{\d H_J}{\d \mathbf p_a}, $$\be
\dot{\bp}_a= \dot{\la}\bp_{a,\la}+\dot{\k}\bp_{a\,k}=\{\bp_a,h_J\}=-
\f{\d H_J}{\d \bu_a}, \l{3.12}\ee where (\r{3.18}) and then (\r{3.7})
have been used step by step. Therefore, $\bu_{a}$ is the solution of
the sourceless Lagrange equation (\r{2.25}) and $\bp_a= \dot{\bu}_a$.
Q.E.D.

Proposition 3 means that $\a$ in (\r{3.2}) is the coset space index
and the condition (\r{3.5}) is satisfied.

\subsection{\it An example of coset space: scalar theory\l{sec3.3}}

Let us start from conformal $\vp^4$ theory. The exact $O(4)\t O(2)$
invariant solution for this theory is known \C{deAlfaro, actorr}: \be
u(\bx,t)=\le\{\f{-(\ga- \ga^*)^2}
{(x-\ga)^2(x-\ga^*)^2}\ri\}^{1/2}.\l{2.3a}\ee The "Hamiltonian" looks
as follows: \be H(u,p)=\int d\bx \le(\f{1}{2}p^2-\f{1}{2} (\pa_i u)^2
+\f{1}{4}gu^4\ri).\l{}\ee The time-like complex vector \be
\ga_\mu=\x_\mu+i\q_\mu,\quad \ga_\mu\ga^\mu= \ga_0^2-\ga_i^2,\:
i=1,2,3\l{}\ee and $\x_\mu$ and $\q_\mu$ are the real numbers. The
parameters $\{\xi,\eta\}$ form the coset space $W$. Their physical
domain is defined by inequalities: \be -\infty\leq\x_ \mu \leq+
\infty,\quad -\infty\leq\eta_i \leq+ \infty,\quad \q_\mu\q^\mu\geq0.
\l{}\ee

The solution (\r{2.3a}) is regular in the Minkowski metric for
$\q=\sqrt{\q_\mu\q^\mu}\ge 0$ and has the pole singularity at \be
(x-\x)^2=0 \l{53}\ee if $\q=0$. We will regularize it continuing on
the Mills complex-time contour \C{milss}. The solution (\r{2.3a}) has
the finite energy and no topological charge. There also exist its
elliptic generalizations of (\r{2.3a})\C{actorr}.

Let us consider now $(\x,\q)$ as the dynamical variables: \be
(\x,\q)= (\x,\q)(\bx,t) \l{54}\ee assuming that the solution of
equations: $$ \{u,h_J\}-\f{\d H_J}{\d p}=0,~~\{p,h_J\}+ \f{\d H_J}{\d
u}=0, $$ see (\r{3.7}) and $$ \dot{\x}(\bx,t)-\frac{\d h_J(\x,\q)}{\d
\q(\bx,t)}=0,~~\dot{\q}_\a(\bx,t)+\f{\d h_J(\x,\q)}{\d\x(\bx,t)}=0,
$$see (\r{3.8}), where
$$h_J(\x,\q)= H_J(u,p), $$ coincides with (\r{2.3a}) in the classical
limit $J=0$. In that limit we have the equations: \be
\sum_\a\le(\f{\pa u}{\pa\x_\a(\bx,t)}\f{\d h}{\d\q_\a(\bx,t)}-\f{\pa
u}{\pa\q_\a(\bx,t)}\f{\d h}{\d\x_\a(\bx,t)}\ri)=\f{\d H}{\d
p(\bx;\x,\q)}=p(\bx;\x,\q),$$$$\sum_\a\le(\f{\pa
p}{\pa\x_\a(\bx,t)}\f{\d h}{\d\q_\a(\bx,t)}-\f{\pa
p}{\pa\q_\a(\bx,t)}\f{\d h}{\d\x_\a(\bx,t)}\ri)=-\f{\d H}{\d
u(\bx;\x,\q)}=-( \pa_i^2u + g u^3)(\bx;\x,\q) \l{55}\ee and \be
\dot{\x}_\a(\bx,t)-\frac{\d h(\x,\q)}{\d \q_\a(\bx,t)}=0,~~
\dot{\q}_\a(\bx,t)+\f{\d h(\x,\q)}{\d\x_\a(\bx,t)}=0. \l{56}\ee for
$u(\bx;\x,\q)$ and $p(\bx;\x,\q)$. First equation in (\r{55}) means
that $$ \{u(\bx;\x,\q),u(\bx;\x,\q)\}\f{\d H}{\d
u(\bx,t)}+\le(\{u(\bx;\x,\q),p(\bx;\x,\q)\}-1\ri)\f{\d H}{\d
p(\bx,t)}=0. $$ Therefore, $u$ and $p$ must obey the equation: \be
\sum_\a\le(\f{\pa u(\bx;\x,\q)}{\pa\x_\a(\bx,t)}\f{\d
p(\bx;\x,\q)}{\d\q_\a(\bx,t)}-\f{\pa u(\bx;\x,\q)}{\pa\q_\a(\bx,t)}
\f{\d p(\bx;\x,\q)}{\d\x_\a(\bx,t)}\ri)=\{u(\bx;\x,\q),p(\bx;\x,\q)\}
=1.\l{58a}\ee But this is impossible solution since $u$ and $p$ are
not the canonically conjugated coordinate and momentum in the
considered theory with symmetry. It must be noted that (\r{58a}) is
the unique consequence of Eqs. (\r{55}).

One can consider another variables: \be \x=\x(t),~~\q=\q(t).
\l{59}\ee In this case one can find: \be \int
d\by\{u(\bx;\x,\q),u(\by;\x,\q)\}\f{\d H}{\d u(\by,t)}+\int
d\by\le(\{u(\bx;\x,\q),p(\by;\x,\q)\}-\d(\bx-\by)\ri)\f{\d H}{\d
p(\by,t)}=0.\l{60}\ee One may consider following solution of this
equation:
$$\{u(\bx;\x,\q),u(\by;\x,\q)\}=0,~~\{u(\bx;\x,\q),p(\by;\x,\q)\}=
\d(\bx-\by)$$ which is equivalent of (\r{58a}) and must be rejected
if ${\d H}/{\d u(\by,t)}$ and ${\d H}/{\d p(\by,t)}$ are not the
independent quantities. Therefore Eq. (\r{60}) must be satisfied only
in the integral over the 3-coordinate sense. Just that case is
realised.

Let us assume now that the variables $\q_\a$ and $\x_a$ are chosen so
that \be h=h(\q).\l{61}\ee In this case Eqs. (\r{56}) looks as
follows: \be \dot{\x}_\a(t)=\frac{\d h(\q)}{\d
\q_\a(t)}\equiv\o_\a(\q),~~ \dot{\q}_\a(t)=-\f{\d
h(\q)}{\d\x_\a(t)}=0 \l{62}\ee i.e. in the considered semi-classical
approximation $\q_\a$ are the integrals of motion and ${\d
h(\q)}/{\d\q_\a(t)}=\o_a(\q)$ are constant velocities in the factor
space $W$, i.e. \be
{\x}_\a(t)=\o_a(\q^0)(t-t^0),~~{\q}_\a(t)={\q}_\a^0,\l{62}\ee where
$t^0$ and ${\q}_\a^0$ are the time independent constants.

\subsection{\it An example: Non-Abelian gauge theory\l{sec.3.4}}

The Gribov ambiguity actually presents the problem in the non-Abelian
gauge theory since we know, at least, the $O(4)\t O(2)$-invariant
strict solution of the $SU(2)$ Yang-Mills equation. The corresponding
coset space has $\dim W=8$ plus the (infinite) gauge groups
dimension.

The $ansatz$ \C{t'hooft-1, uy}: \be \sqrt{g}u_\mu^a= \q^a_{\mu\nu}
\pa^\nu\ln u\l{2.2}\ee  for the $SU(2)$ Yang-Mills potential
$u_\mu^a$ leads to the conformal scalar, $\vp^4$, field theory, see
e.g. \C{actorr}, which was considered in previous subsection.

\section{Reduction\label{re1}}

After having done the mapping, see (\r{3.20})-(\r{3.22}), one must
extract from the set $\{\k,\la\}$ the dynamical variables
$\{\xi,\eta\}$.

\subsection{\it Cyclic variables}

Let us divide the set $\{\la,\k\}$ into two parts:
\be\{\la,\k\}\to(\{\la,\k\},\{\x',\q'\}),\l{3.19a}\ee assuming
that $\la$ and $\k$ are cyclic variables: \be
\f{\pa\bu_a}{\pa\la}\approx 0,~~\f{\pa\bu_a}{\pa\k}\approx
0\l{3.21}\ee and the derivatives of $\bu_a$ over $\x'$ and $\q'$
not vanish at $\vep=0$. It can be shown that the variables
$(\la,\k)$ stay cyclic in the quantum sense as well.

\vskip0.25cm {\bf Proposition 4. \it The quantum force is orthogonal
to the cyclic variables axes.}\\ Indeed, taking into account
(\r{3.21}), \be \bk(j,e)=\int dt \le\{\f{\d}{\d j_{\la}}\cdot
\f{\d}{\d e_{\la}}+ \f{\d}{\d j_{\k}}\cdot \f{\d}{\d
e_{\k}}+\f{\d}{\d j_{\x'}}\cdot \f{\d}{\d e_{\x'}}+ \f{\d}{\d
j_{\q'}}\cdot \f{\d}{\d e_{\q'}}\ri\}.\l{4.3} \ee As it follows from
(\r{3.18}), \be \f{\d \bu_a}{\d j_X}\sim\f{\d\bu_a}{\d X}\approx
0,~X=(\la,\k).\l{58}\ee Therefore, we can write taking into account
(\r{3.21}) and (\r{58}) that \be 2\bk(j,e)=\int dt \le\{\f{\d}{\d
j_{\x'}}\cdot \f{\d}{\d e_{\x'}}+ \f{\d}{\d j_{\q'}}\cdot \f{\d}{\d
e_{\q'}}\ri\}.\l{4.5}\ee Then, following our definition, one should
take everywhere \be j_X=e_X=0,~X=(\la,\k).\l{}\ee

\vskip 0.25cm The result of the reduction looks as follows:\be
DM(u,p)=d\O~ DM(\x',\q'),\l{4.7}\ee where the infinite dimensional
integral over \be d\O=\prod_{\a,t} d\la_\a(t)
d\k_\a(t)\d(\dot\la_\a(t)) \d(\dot\k_\a(t))\l{4.8}\ee will be
cancelled by normalization. This procedure completes the
renormalization in the transformed formalism.

The remaining degrees of freedom are entered into the reduced Dirac
measure: \be DM(\x',\q')=\prod_t d\x'(t)d\q'(t) \:
\d(\dot\x'-h_{\q'}(\x',\q')-j_{\x'}) \d(\dot\q'+
h_{\x'}(\x',\q')-j_{\q'}).\l{7a} \ee This result presents the first
step of the reduction into the physical coset space $W$.  Q.E.D.

Let us consider now the case when only the part of variables are
cyclic: $\{\x'\}=(\{\x\},\{\x''\})$ and
$\{\q'\}=(\{\q\},\{\q''\})$,
$$\dim\{\x\}=\dim(\{\q\})$$
where, for example, only $\{\x''\}$ is the set of cyclic variables
(the case when $\{\q''\}$ is cyclic is similar): \be
\f{\pa\bu_a}{\pa\x''}\approx0. \l{4.9}\ee

It can be easily show that we come to the condition of Proposition
4, and in this case the conjugated variables $\{\q''\}$ are the integrals of motion.

Indeed, in the frame of the definition (\r{4.9}) we have
\be
DM(\x,\q;\q'')=\prod_t
d\x(t)d\q(t)d\x''(t)d\q''(t)
\d(\dot\x-h_{\q}-j_{\x})\d(\dot\q+h_\x-j_{\q})\d(\dot\q''-j_{\q''}).
\l{4.10}
\ee

Following (\r{3.22}) the virtual deviation $\bE$ looks as follows:
\be \bE_a= e_{\q}\bu_{a\x}- e_{\x}\bu_{a\q}
+e_{\q''}\bu_{a\x''}\l{e13}\ee and the perturbations generating
operator is: \be 2\bk(j,e)=\int dt \le\{ \f{\d}{\d
j_{\x}}\f{\d}{\d e_{\x}}+ \f{\d}{\d j_{\q}}\f{\d}{\d e_{\q}}+
\f{\d}{\d j_{\q''}} \f{\d}{\d e_{\q''}}\ri \}.\l{4.12a}\ee As it
follows from the general condition that the auxiliary variables
must be taken equal to zero, we must put $e_{\q''}=0$ because of
(\r{4.9}), (\r{e13}). We must omit simultaneously the last term in
(\r{4.12a}). For this reason one must put $j_{\q''}=0$ in
(\r{4.10}) and therefore $\q_\a''$ are the integrals of motion.

\vskip 0.25cm Following to this section one can conclude that
gauge degrees of freedom can not belong to the quantum variables,
$\{\La_a\}\nsubseteq\{\x,\q\}$, since there is no conjugated to
$\La^a$ gauge charge dependence in the field $u_{a\mu}$.

\subsection{\it Concluding expression}

As a result, \be \cn=e^{-i\mathbf{k}(je)} \int DM(\x,\q)
e^{iU(\bu,\bE)}, \l{18}\ee where the new coset space virtual
deviation is \be \bE_{a}=\sum_\a\le\{
e_{\q_\a}\f{\pa \bu_{a}}{\pa\x_\a}-e_{\x_\a}\f{\pa \bu_{a}}
{\pa\q_\a}\ri\}.\l{20}\ee

The generating quantum perturbations operator in the coset space
is \be 2\mathbf{k}(je)=\sum_\a\int dt\le\{\f{\d}{\d
j_{\x_\a}(t)}\f{\d}{\d e_{\x_\a}(t)}+\f{\d}{\d j_{\q_\a}(t)}
\f{\d}{\d e_{\q_\a}(t)}\ri\}, \l{4.11}\ee where summation is
performed over all canonical pairs, $(\x,\q)\in T^*W$. Let us choose
the variables $\{\x,\q\}$ so that $\pa h/\pa\x=0$, then the
corresponding measure is
\be
DM=dR\prod_{\a,t}
d\x_\a(t)d\q_\a(t) \:\d\le(\dot\x_\a- h_{\q_\a}(\q)-j_{\x_\a}\ri)
\d\le(\dot\q_\a-j_{\q_\a}\ri), \l{4.12}
\ee
where  $dR$ is the zero
modes Cauchy measure: \be dR=\prod_{\a,t} d\q_\a''\d(\dot\q_\a'').
\l{4.13}\ee Therefore, \be W=T^*W+R\l{}\ee where $\{\x,\q\}\in
T^*W$ and $\{\x''\}\in R$.

The coset space Hamiltonian equations: \be \dot\x_\a-
h_{\q_\a}(\q)=j_{\x_\a},~\dot\q_\a=j_{\q_\a}\l{4.22}\ee are easily
solved through the Green function $g(t-t')$. The latter must obey
the equation: \be \pa_tg(t-t')=\d(t-t').\l{4.23}\ee This Green
function has the universal meaning, and it must be the same for
arbitrary theory. Then, using the $i\vep$-prescription and the
experience of the Coulomb problem considered in \C{jmp-1}, we will
use the following solution of (\r{4.23}):\be g(t)=\le\{
\begin{array}{c}
  1,~~ t\geq0 \\
  0,~~ t<0
\end{array}.\ri.\l{}\ee The solution of the Eq.(\r{4.22}) looks
as follows: \be \x_a^j(t)=\int dt' g(t-t')\{h_{\q_\a}(\q^j)+
j_{\x_\a}\}(t'),\ee \be\q_a^j(t)=\int dt g(t-t')j_{\q_\a}(t').
\l{}\ee As a result, the functional measure $DM$ is reduced to the
Cauchy measure \be dM=\prod_{\a,t} d\q_\a''(t)\d(\dot\q_\a'')
d\x_\a(t)d\q_\a(t)\d(\dot\x_\a) \d(\dot\q_\a) = \prod_{\a}
d\q_\a''(0)d\x_\a(0)d\q_\a(0).\l{} \ee The integral over $dM$
gives the volume ${V}$ of the factor group $\cg/\ch$ and
$\dim{V}\leq\dim W$.

Notice that the gauge group volume ${V}_\La$ in our formalism is
defined by the measure $\prod_{a,\bx} d\La_a(\bx,0)$.

Therefore, \be \cn=e^{-i\mathbf{k}(je)} \int dM
e^{iU(\bu^j,\bE^j)}, \l{4.27}\ee where $\bu^j$ and $\bE^j$ depends
on the functions $(\x_a^j, \q_a^j)$.

\section{\it Gauge invariance\label{g:inv}}

The new coset space virtual deviation  \be \bE_{a}=\sum_\a\le\{
e_{\q_\a}\f{\pa \bu_{a}}{\pa\x_\a}-e_{\x_\a}\f{\pa \bu_{a}}
{\pa\q_\a}\ri\}.\ee is the covariant of gauge transformations: if
\be u_{a\mu} \to \O u_{a\mu}\O^{-1}+i\:\O \pa_\mu\O^{-1} \l{g:u}
\ee then \be \bE_{a}\to \O \bE_{a}\O^{-1}\l{2.14},\ee because of
the condition $\{\La_a\}\nsubseteq\{\x,\q\}$.

Following (\r{2.17a}) and (\r{20}),
\begin{multline}
U(u,e)=S(u+e)-S^*(u-e) -2\Re
\sum_{a,\a} \int_C dx \le\{
e_{\q_\a} \f{\pa \ru_{ai}(x)}{\pa\x_\a}-e_{\x_\a}\f{\pa \ru_{ai}(x)}{\pa\q_\a}
\ri\}\f{\d S(u)}{\d
u_{ai}(x)} \\ = S(u+e)-S^*(u-e)-2\Re\sum_{\a}\int_C dt
\le\{e_{\q_\a}(t)\f{\d}{\d\x_\a(t)}-e_{\x_\a}(t)
\f{\d}{\d\q_\a(t)}
\ri\}S(u).\l{}
\end{multline}
This quantity is transparently gauge invariant, since the action
$S$ is the invariant of gauge transformation (\r{g:u}),
(\r{2.14}).

We can conclude that each term of the coset space perturbation theory
is gauge invariant since $DM$ in (\r{4.12}) and $\bk(je)$ in
(\r{4.11}) are the gauge invariant quantities.

It is interesting to note that in spite of the fact that each term
of the perturbation theory is transparently gauge invariant,
nevertheless, one can not formulate the theory only in terms of
the gauge field strength.

\section{Conclusions}

It is useful to summarize the rules of the coset space perturbation
theory.

(i) The transformation, (\r{1.2}), to independent variables is
performed having in mind that the power of the variables set must not
be altered.

(ii) The "host free" transformation is induced by the function
$\bu_a$ defined by the Eq.(\r{3.7}). In this stage the function
$h_j(\la,\k)$ is arbitrary.

(iii) If $h_J$ is is the transformed Hamiltonian, $h_J=h+\int
dx\:\bu_a \mj_a$, then there exists a mapping into the $W$ space, see
(\r{1.2}). This mapping produces a new set of sources
$\{j_\la,j_\k\}$, (\r{3.15b}), and virtual deviations,
$\{e_\la,e_\k\}$, (\r{3.22}). It is remarkable that each degree of
freedom of the $(\la,\k)$ space is excited independently of one
another by the individual sources $\{j_\la,j_\k\}$. This is crucial
for the reduction of the quantum degrees of freedom.

(iv) One can consider the case when a subset of variables is
cyclic, see (\r{3.19a}) and (\r{3.21}). As a result we have found
the reduced measure (\r{7a}), and the perturbations generating
operator (\r{4.5}). The volume of the cyclic variables, (\r{4.8}),
is cancelled by normalization. The field theoretical problem
becomes finite dimensional. The cancellation of the cyclic
variables volume can be considered as a renormalization procedure.

(v) $W$ is the coset space. The
choice of the coset variables $\{\x,\q\}$ is arbitrary.

(vi) A portion of the remaining variables can belong to the
symplectic subspace $T^*W\subseteq W$. The latter allows to conclude that the gauge phase
$\La_a$ can not belong to $T^*W$. As a result the perturbation
theory is transparently gauge invariant.

(vii) The known solution \C{deAlfaro} shows that all space-time
integrals of the coset space perturbation theory are finite
outside the border $\pa W$ since $|S(u)|<\infty$ and $\dim W$ is
finite. The border contributions, $\sup(\x,\q)\in \pa W$, remain
finite because of the $i\vep$-prescription. Further analysis of
the role of the border singularities, see also \C{jmp-1}, will be
given in subsequent publications.

\begin{acknowledgements}
We would like to thank V. Kadyshevsky and our colleagues in the
Lab. of Theor. Phys. of JINR for their deep interest in the
described approach.

This work was supported by Russian Foundation for Basic Research
($\#$ 05-02-17693).
\end{acknowledgements}

\theendnotes

\end{document}